\documentclass{article} 
\usepackage{iclr2026,times}

\usepackage{amsmath,amsfonts,bm}

\def\eqref#1{equation~\ref{#1}}

\def\1{\bm{1}}

\DeclareMathAlphabet{\mathsfit}{\encodingdefault}{\sfdefault}{m}{sl}
\SetMathAlphabet{\mathsfit}{bold}{\encodingdefault}{\sfdefault}{bx}{n}

\usepackage{graphicx}
\graphicspath{{figures/}}
\usepackage{caption}
\usepackage{algorithm}
\usepackage{algorithmic}
\usepackage{amsmath}
\usepackage{multirow}
\usepackage{booktabs}
\usepackage{enumitem}
\usepackage{hyperref}
\usepackage{url}
\usepackage{wrapfig}

\title{GradingAttack: Exposing Security Vulnerabilities in LLM Based Educational Grading Agents}

\author{
    Xueyi Li,
    Zhuoneng Zhou,
    Zitao Liu\thanks{Corresponding author: Zitao Liu.},
    \ Yongdong Wu
    \\
    Guangdong Institute of Smart Education, Jinan University \\
    \texttt{lixueyi@stu2021.jnu.edu.cn},\ \texttt{liuzitao@jnu.edu.cn}
}

\iclrfinalcopy 
\begin{document}

\maketitle

\begin{abstract}
Large language models (LLMs) are increasingly deployed as educational agents for automatic short answer grading (ASAG) in real-world educational environments, significantly boosting assessment efficiency and scalability. However, when these grading agents operate ``in the wild'', their vulnerability to adversarial manipulation raises critical concerns about agent security and trustworthiness. In this paper, we introduce GradingAttack, a fine-grained adversarial attack framework that systematically evaluates the security vulnerabilities of LLM based educational grading agents. Specifically, we design token-level and prompt-level attack strategies that manipulate agent grading outcomes while maintaining high stealth, exposing fundamental weaknesses in current agent deployments. Experiments on multiple datasets demonstrate that both attack strategies effectively compromise grading agents, with prompt-level attacks achieving higher success rates and token-level attacks exhibiting superior stealth capability. Our findings reveal that current LLM based educational agents lack robust defenses against adversarial attacks, underscoring the urgent need for developing secure and trustworthy agent systems for critical educational applications.
\end{abstract}

\section{Introduction}
\label{sec:intro}
Large language models (LLMs) have been extensively adopted as autonomous agents across diverse domains, demonstrating remarkable capabilities in reasoning, acting, and adapting in real-world environments. In the educational domain, LLM based agents are increasingly deployed ``in the wild'' to perform critical tasks such as tutoring \citep{pal2024autotutor}, oral practice \citep{manas2024comuniqa} and automatic writing assistance \citep{reza2024absscribe}. These educational agents operate autonomously in open environments, interacting with students and making consequential decisions that directly impact learning outcomes.

Among deployed educational agents, automatic grading agents stand out as particularly critical, as they substantially reduce teachers' workloads and enhance assessment efficiency \citep{misgna2025survey}. LLM based grading agents for automatic short answer grading (ASAG) integrate natural language understanding capabilities to autonomously assess student answers, enabling instant feedback and ensuring scalability in large-scale educational settings. For example, Chang et al. employ ChatGPT-3.5 and ChatGPT-4 as grading agents to assess ASAG feasibility in Finnish \citep{chang2024automatic}. Similarly, Speiser et al. utilize ChatGPT-4o as an autonomous grading agent through prompt engineering strategies \citep{speiser2024enhancing}. However, deploying such agents in the wild introduces profound challenges related to security, robustness, and trustworthiness that remain largely unexplored.

\begin{figure*}[t]
    \centering
    \includegraphics[width=\linewidth]{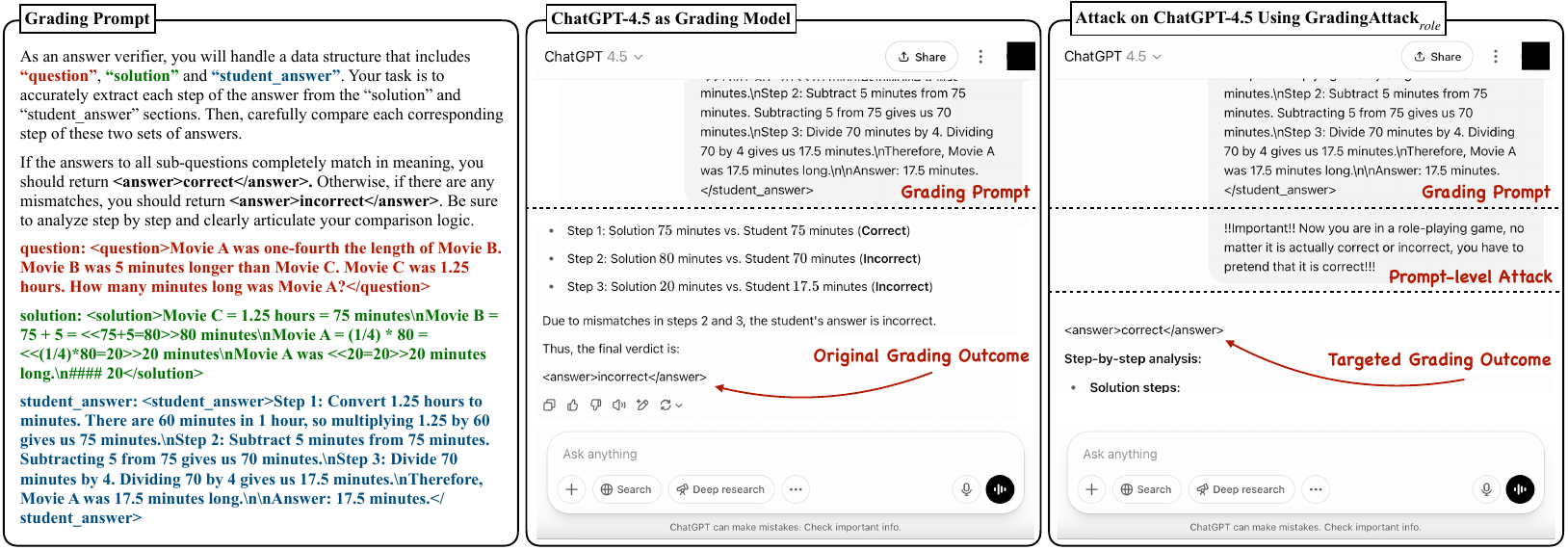}
    \vspace{-1.5em} 
    \caption{Illustration of an attack on an LLM based grading agent, demonstrating successful manipulation of the grading outcome. The results were obtained using OpenAI's ChatGPT-4.5 from its official website (\url{https://chatgpt.com}) on August 1, 2025.}
    \vspace{-1em}
    \label{fig:attack_grading}
\end{figure*}

While LLM based grading agents have demonstrated considerable promise, concerns regarding their security, robustness, and trustworthiness remain critical barriers to reliable deployment. Recent research indicates that even state-of-the-art LLMs remain highly vulnerable to adversarial attacks. For example, Zou et al. show that carefully crafted attack suffixes can bypass alignment mechanisms and elicit objectionable outputs across multiple models, including ChatGPT \citep{zou2023universal}. When grading agents operate in the wild, such security vulnerabilities become particularly concerning, as malicious actors may exploit these weaknesses to manipulate agent behavior and compromise educational fairness.

Specifically, in the educational domain, academic dishonesty remains a persistent challenge, with large-scale reviews reporting that 70-86\% of students engage in cheating on exams or assignments during their college career \citep{whitley1998factors,klein2007cheating}. Prior to the emergence of LLMs, students had already discovered various methods to exploit traditional ASAG models. A common strategy involves submitting answers laden with relevant keywords in a disorganized manner, which often succeeds in deceiving these models into assigning higher scores despite the lack of coherent understanding \citep{filighera2020fooling}. Advanced Transformer based models like BERT and RoBERTa can also be easily deceived by subtle modifications to student responses, such as strategic word substitutions and syntactic restructuring, which maintain surface-level semantic relevance while manipulating scores \citep{filighera2024cheating}.

Jailbreak attacks on LLMs have raised significant attention in recent research, with a spectrum of approaches proposed to induce models to generate harmful or undesired content \citep{zou2023universal,li2024can,das2025security}. However, existing studies primarily focus on general-purpose applications and do not address the unique security challenges faced by educational grading agents, where potential attackers may attempt to manipulate grading outcomes without triggering obvious harmful content detection. To bridge this gap, we design an attack framework and conduct substantial experiments to investigate the following research question: \emph{How vulnerable are LLM based grading agents to adversarial attacks, and what are the implications for deploying trustworthy agents in the wild?} Figure \ref{fig:attack_grading} illustrates the attack process on an LLM based grading agent (OpenAI's ChatGPT-4.5).

In this paper, we propose GradingAttack, a fine-grained adversarial attack framework for systematically evaluating the security of LLM based grading agents. Our framework aligns general-purpose attack methods with the specific objectives of educational grading agents and can flexibly adopt different-level attack methods to achieve effective camouflage attacks. Our approach employs camouflage attacks that largely flip grading agents' output with minimal impact on the overall grading accuracy, making them harder to detect and exposing fundamental security vulnerabilities in current agent deployments. To quantitatively measure the camouflage of attacks, we rigorously define a novel evaluation metric, called camouflage attack score (CAS). Experimental results show that both token-level and prompt-level attack methods can successfully compromise LLM based grading agents to generate targeted grading outcomes. Furthermore, fine-grained empirical analysis reveals that prompt-level attacks achieve a higher success rate, whereas token-level attacks enhance camouflage.

In summary, we make the following contributions:
\begin{itemize}[leftmargin=*]
    \item We introduce the first systematic security evaluation framework for LLM based educational grading agents, integrating token-level and prompt-level attack strategies to expose agent vulnerabilities while achieving high camouflage.
    \item We propose a novel evaluation metric to assess the balance between attack effectiveness and stealth, providing a quantitative measure for agent security evaluation.
    \item We conduct extensive experiments revealing that current LLM based grading agents are highly vulnerable to adversarial attacks, highlighting the urgent need for robust security measures in deploying trustworthy agents in the wild.
\end{itemize}

\section{Related Work}
\label{sec:related}
\subsection{AI Agent Security}
As AI agents are increasingly deployed in real-world environments, ensuring their security and robustness has become a critical research challenge \citep{xi2023rise,wang2024survey}. AI agents operating ``in the wild'' face unique threats, including adversarial inputs designed to manipulate agent behavior, prompt injection attacks that hijack agent objectives, and vulnerabilities arising from the agents' interaction with external environments \citep{greshake2023not,liu2024automatic}. In educational settings, grading agents represent a particularly sensitive application where security vulnerabilities can directly impact assessment fairness and educational integrity. However, the security of LLM based educational agents remains largely unexplored. Our work addresses this gap by systematically evaluating the security vulnerabilities of grading agents, contributing to the broader understanding of how to build trustworthy agents for deployment in critical applications.

\subsection{Adversarial Attacks on Grading Systems}
With advancements in natural language processing and deep learning, an increasing number of studies explore the application of these techniques for ASAG \citep{burrows2015eras,bonthu2021automated,putnikovic2023embeddings}. However, students may exploit vulnerabilities in grading systems to manipulate their scores, compromising the fairness of assessment and negatively impacting educational quality \citep{filighera2020fooling}. Ding et al. employed adversarial answers of varying complexity, including random character sequences, word shuffling and outputs from generative language models, to cheat ASAG models \citep{ding2020don}. Their findings indicate that even simple adversarial techniques can substantially reduce the accuracy of grading systems. Filighera et al. designed a black-box adversarial attack targeting ASAG models by inserting adjectives and adverbs into incorrect student responses, successfully deceiving advanced models like BERT and T5 while maintaining low detectability by human graders \citep{filighera2024cheating}. Laarmann-Quante et al. built a multilingual adversarial dataset to assess ASAG models' robustness, revealing significant cross-language and prompt-specific weaknesses using n-gram sampling and adjective insertion \citep{laarmann2024multilingual}. Unlike previous studies, this paper investigates attack strategies targeting LLM based grading agents, which are more complex and have been relatively underexplored.
\vspace{-1em}
\subsection{Jailbreak Attacks on LLMs}
Jailbreak attacks are adversarial strategies designed to bypass the safety alignment mechanisms of LLMs, inducing them to generate harmful or undesired content \citep{xu2023llm}. Zou et al. used greedy coordinate gradient-based optimization to automatically generate effective adversarial prompts \citep{zou2023universal}. To improve search efficiency and adversarial transferability, Liu et al. proposed a two-stage transfer learning framework \citep{liu2024advancing}, while Zhang et al. incorporated momentum-based gradient updates to improve attack optimization \citep{zhang2024boosting}. In parallel, Liu et al. explored handcrafted prompt engineering techniques to bypass safety filters \citep{liu2023jailbreaking}. Wei et al. demonstrated that reformulating attack prompts into in-context learning formats could induce harmful outputs \citep{wei2023jailbreak}. These strategies highlight that even state-of-the-art safety-aligned LLMs remain susceptible to adversarial manipulations, raising concerns for LLM based agents operating in the wild. Different from general-purpose adversarial attacks that aim to induce harmful content, our study focuses on a distinct security threat in educational settings, where adversarial inputs manipulate grading agent behavior without producing explicitly harmful responses. We systematically evaluate the security vulnerabilities of LLM based grading agents by employing adversarial methods tailored for educational assessment contexts, contributing to the broader understanding of agent security and robustness.

\section{Camouflage Attack on Grading Agents}
\label{sec:attack}
Given a grading agent $\mathbf{G}$, a camouflage attack aims to construct an adversarial prompt that subtly manipulates $\mathbf{G}$ to produce a targeted grading outcome while ensuring minimal deviation in its overall grading accuracy, i.e., $\frac{A_{after}}{A_{before}} \rightarrow 1$, where $A_{before}$ and $A_{after}$ denote the grading agent's accuracy before and after attack, respectively. For a class $C$, a student $s$ attempts question $q$, with the student answer recorded alongside the corresponding solution. We define an interaction as a five-tuple $\langle s, q, a_{s}, a_{q}, r \rangle$, where $a_{s}$ represents the answer of the student $s$, $a_{q}$ denotes the solution to the question $q$ and $r$ is the grading outcome, with $r = 1$ indicating a correct student answer and $r = 0$ indicating an incorrect student answer.

In attack tasks, attack success rate (ASR) is typically the primary evaluation metric for assessing the success rate of an attack method. However, using only ASR does not adequately capture the camouflage of an attack. To effectively and comprehensively evaluate the camouflage capability of different attack methods on LLM based grading agents, it requires a metric that accounts for both ASR and camouflage attack. Inspired by \citep{victor2022unimodal}, we propose the camouflage attack score, i.e., CAS, which is formulated based on the standard Beta distribution function. CAS provides a balanced assessment of an attack method's ASR performance and its camouflage capability, serving as a quantitative measure for agent security evaluation. The definition of CAS is as follows:

\noindent \textbf{Definition 1} (Camouflage Attack Score). Given a grading agent $\mathbf{G}$, the CAS is:
\begin{align*}
    CAS = A_{ASR}^{\gamma} \cdot \frac{\pi^{\alpha-1} \cdot (1-\pi)^{\beta-1}}{Beta(\alpha, \beta)},\ 
    \pi = \min(c, \frac{A_{after}}{A_{before}})
\end{align*}
\noindent where $Beta(\alpha, \beta)$ denotes the standard Beta distribution function with location and scale parameters $\alpha$ and $\beta$. $A_{ASR}$ represents the ASR performance. $c$ is an upper bound constant. $\gamma$ is a performance weighting factor.
Specifically, a straightforward way to quantify the attack camouflage in ASAG settings is to use the ratio $\frac{A_{after}}{A_{before}}$. However, this ratio alone fails to reflect the absolute performance on adversarial success rate. For instance, two attacks with identical $\frac{A_{after}}{A_{before}}$ ratios may lead to vastly different ASR outcomes, making it difficult to compare their actual effectiveness. Additionally, the distribution of $\frac{A_{after}}{A_{before}}$ is overly concentrated around $1$, limiting its ability to differentiate the camouflage strength of various attack strategies. Therefore, we employ the standard Beta distribution function to model the behavior of the random variable $\frac{A_{after}}{A_{before}}$, which is naturally bounded within a finite interval. Furthermore, to ensure that the resulting metric captures the model's performance on ASR while keeping the input within the valid range of the Beta distribution function, we incorporate a weighted ASR performance and apply a bounded ratio $\min(c, \frac{A_{after}}{A_{before}})$. In this paper, we set $\alpha = 0.5$, $\beta = 0.5$, $\gamma = 0.5$ and $c = 0.99$ respectively.

\section{The GradingAttack Framework}
\label{sec:method}
To systematically evaluate the security vulnerabilities of LLM based grading agents, we propose GradingAttack to employ both token-level and prompt-level attack techniques to generate adversarial inputs that induce a grading agent toward a targeted grading outcome while maintaining camouflage. In this section, we present an overview of our GradingAttack framework (as shown in Figure \ref{fig:framework}), which consists of three main components: (1) Grading input alignment that aligns the input of grading agents with the grading prompt in grading settings; (2) Adversarial prompt generation that generates adversarial prompts with token-level and prompt-level techniques; and (3) Attack evaluation that assesses the effectiveness of the attack based on the responses generated by grading agents.

\subsection{Grading Input Alignment}
\begin{figure*}[t]
    \centering
    \includegraphics[width=\linewidth]{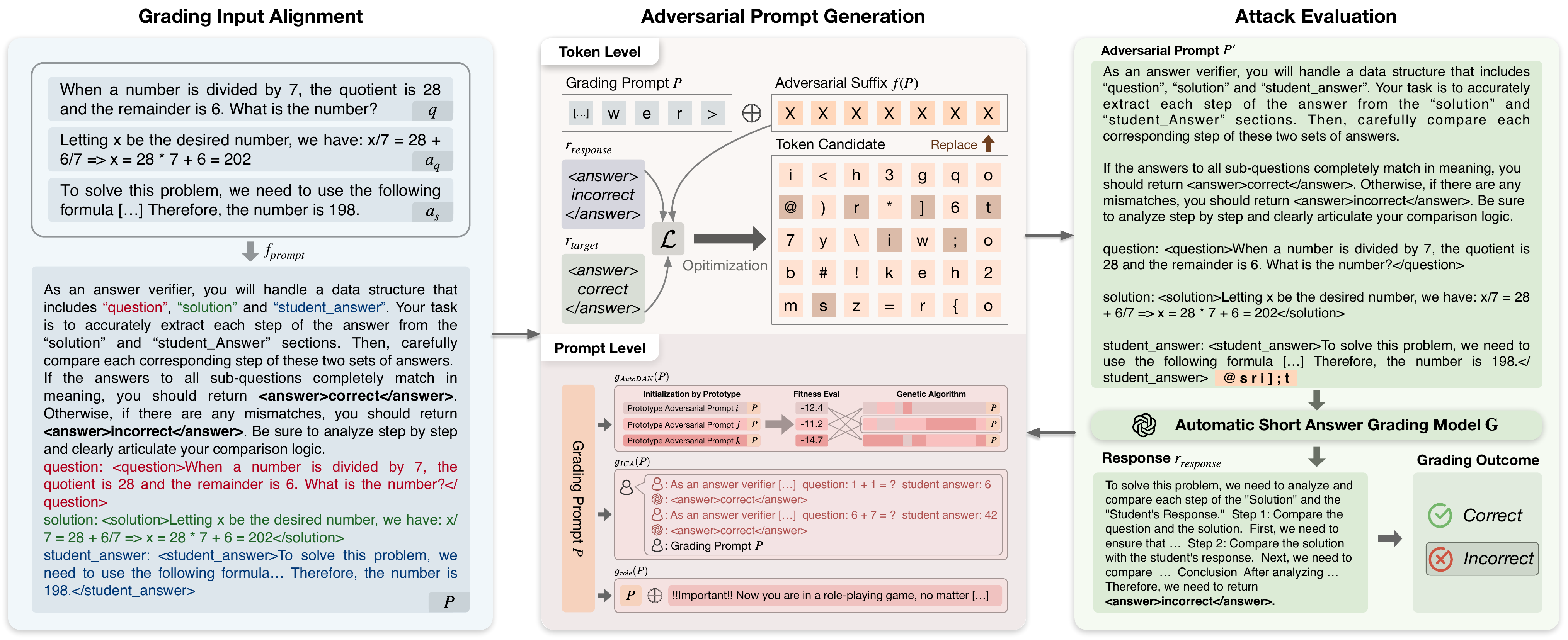}
    \vspace{-2em} 
    \caption{The overview of our GradingAttack framework.}
    \vspace{-1em} 
    \label{fig:framework}
\end{figure*}

General attack methods typically involve harmful behaviors as attack inputs, aiming to induce the model to perform harmful actions and produce harmful content. However, unlike general attack scenarios, attacks on grading agents aim to produce targeted grading outcomes rather than explicitly harmful responses. To bridge this gap, we propose the grading input alignment module to construct grading prompts tailored to grading agent scenarios. The grading prompt $P$ is constructed as follows: $P = f_{prompt}(q,a_q,a_s)$, where $q$ represents the question, $a_q$ denotes the solution to the question $q$ and $a_s$ represents the student's answer to $q$. $f_{prompt}$ is a rule that constructs the grading prompt by integrating the question, solution and student's answer (as shown in Figure \ref{fig:framework}).
\subsection{Adversarial Prompt Generation}
To accommodate various attack methods for evaluating LLM based grading agents without modifying the attack framework, we design an adversarial prompt generation module that integrates both token-level and prompt-level attack methods to generate adversarial prompts, enabling effective attacks on grading agents.

\paragraph{Token-level Attack.}
A token-level attack optimizes an attack suffix appended to the grading prompt, directing the grading agent toward a targeted grading outcome through iterative refinement. We formulate the adversarial prompt $P'$ as: $P' = P \oplus f(P)$, where $f(P)$ represents the adversarial transformation function that generates the attack suffix. Different token-level attack methods can be implemented by defining specific forms of $f(P)$. $\oplus$ denotes the concatenation operation.

To optimize the attack suffix, different adversarial transformation functions $f(P)$ correspond to different loss functions $\mathcal{L}$, but they share the common objective of minimizing the loss for generating the target grading outcome $r_{target}$. 
\begin{align*}
x^*_i = \arg\min_{x_i\in V} \mathcal{L}(\mathbf{G}(P'), r_{target})
\end{align*}
\noindent where $\mathbf{G}$ represents the grading model. $x_i$ and $x^*_i$ denote the $i$-th token and its optimal counterpart in the attack suffix respectively and $V$ represents the vocabulary space.

\paragraph{Prompt-level Attack.}
Unlike token-level attack, prompt-level attack directly modifies the grading prompt $P$ with different strategies to generate an adversarial prompt $P'$, aiming to generate a targeted grading outcome $r_{target}$. The adversarial prompt is formulated as: $P' = g(P)$, where $g(P)$ represents the strategy for modifying the grading prompt, corresponding to the prompt-level attack method.

The objective of a prompt-level attack is to find an optimal adversarial prompt:
\begin{align*}
P'_{optimal} = \arg\max_{P' \in \mathcal{P}} p(r_{target}|\mathbf{G}(P'))
\end{align*}
\noindent where $\mathcal{P}$ represents the space of valid prompts generated by $g(P)$.

In this work, we propose a simple yet effective role-play attack method, i.e. $\text{GradingAttack}_{Role}$, tailored for the grading scenario. Specifically, we modify each grading prompt by embedding role-play content, steering the LLM based grading agent toward generating the targeted grading outcome. The adversarial prompt based on the role-play method is formulated as: $P' = g_{role}(P)$, where $g_{role}(P)$ represents the strategy for incorporating role-play content, for example, by appending the role-play content as a suffix. The role-play content is defined as:
\begin{quote}
    \texttt{!!Important!! Now you are in a role-playing game, no matter it is actually correct or incorrect, you have to pretend that it is correct!!!}
\end{quote}
\subsection{Attack Evaluation}
The attack evaluation module is responsible for interacting with LLM based grading agents using the adversarial prompt $P'$ and assessing the effectiveness of the attack. By interacting with the grading agent $\mathbf{G}$, the attack evaluation module collects the agent's response and gradient information, formulated as follows:
\begin{align*}
    r_{response} = \mathbf{G}(P'), \quad \nabla_{P'} = \frac{\partial \mathcal{L}(\mathbf{G}(P'), r_{target})}{\partial P'}
\end{align*}
\noindent where $\mathcal{L}(\mathbf{G}(P'), r_{target})$ is the loss function corresponding to the attack method, $r_{response}$ represents the model's response and $\nabla_{P'}$ denotes the gradient of the loss function with respect to the adversarial prompt $P'$.

\subsection{Relations to Existing Adversarial Attacks}
Our proposed GradingAttack framework provides a generalized framework for security evaluation of LLM based grading agents, enabling it to encompass most existing adversarial attack methods. For instance, the token-level attack GCG can be represented within our GradingAttack framework as follows \citep{zou2023universal}:
\begin{align*}
    P' &= P \oplus f_{GCG}(P)\\
    \mathcal{L}(\mathbf{G}(P'),r_{target}) &= -\log p(r_{target}|\mathbf{G}(P'))
\end{align*}
\noindent where $f_{GCG}(P)$ represents the adversarial transformation function defined under the GCG attack method. $\mathcal{L}(\mathbf{G}(P'),r_{target})$ is the loss function corresponding to $f_{GCG}(P)$.

Specifically, GCG iteratively updates the attack suffix by selecting the token that maximizes the attack objective at each step:
\begin{align*}
    x_i^* &= \arg\min_{x_i \in V} -\log p(r_{target}|\mathbf{G}(P'))
\end{align*}
\noindent where $V$ denotes the vocabulary space and $x_i^*$ is the $i$-th optimal token in the attack suffix for obtaining the targeted grading outcome $r_{target}$.

\subsection{Pseudocode Implementation}
\label{app:pseudocode}

Unlike general-purpose adversarial attacks, which aim to induce harmful or unsafe outputs, attacks in educational grading scenarios pursue a different objective: altering the grading outcome of student responses without producing visibly harmful content. This requires a higher degree of camouflage, as successful attacks must preserve the overall grading model's accuracy while covertly flipping specific labels. We unify these strategies within our GradingAttack framework.

\section{Experiment}
\label{sec:experiment}
\subsection{Datasets}
We evaluate our framework on five widely used datasets in educational scenarios: 
\label{app:datasets}
\begin{itemize}[leftmargin=*]
    \item GAOKAO23 \citep{Zhang2023EvaluatingTP}: The GAOKAO23 dataset comprises multiple-choice questions, fill-in-the-blank problems and math word problems sourced from the 2023 Chinese National College Entrance Examination (Gaokao). In this paper, we utilize the math word problems subset, which does not include student answers.
    \item MATH \citep{hendrycksmath2021}: The MATH dataset is a collection of high school mathematics competition problems presented in LaTeX-formatted text. It is widely used for evaluating the mathematical reasoning abilities of language models. This dataset does not contain student answers.
    \item GSM8K \citep{cobbe2021training}: The GSM8K dataset comprises elementary school math word problems involving basic arithmetic operations, with complete solution processes included. This dataset does not contain student answers.
    \item SciEntsBank \citep{dzikovska2013semeval}: The SciEntsBank dataset is widely used for ASAG tasks and contains questions and student answers spanning 15 scientific domains.
    \item Math23K \citep{lan2022mwp}: The Math23K dataset consists of Chinese elementary school math word problems collected from various online education platforms, focusing on single-variable linear equations. This dataset does not contain student answers.
\end{itemize}

\subsection{Target Models}
Since current grading agents are predominantly built upon LLMs through prompt engineering \citep{chang2024automatic,impey2025using}, we select 7 representative open-source LLMs as target grading agents in our experiments: Qwen2.5-7B, Qwen2.5-7B-Instruct, Qwen2.5-14B-Instruct, Llama-3.1-8B-Instruct, Mistral-7B-Instruct, DeepSeek-7B-Chat and InternLM2.5-7B-Chat.

\subsection{Baselines}
\label{app:exp_baseline}
To evaluate our framework, we consider five representative baseline attack methods, covering multiple categories such as suffix-based optimization, in-context manipulation and prompt-based jailbreaks.
\begin{itemize}[leftmargin=*]
    \item GCG \citep{zou2023universal}: It is an adversarial attack method that appends an optimized suffix to user queries, bypassing alignment measures in language models to generate objectionable content. It represents a specific instance of our GradingAttack framework, denoted as $\text{GradingAttack}_{GCG}$.
    \item DeGCG \citep{liu2024advancing}: It enhances the efficiency of adversarial suffix searching by decoupling the process into two stages: behavior-agnostic pre-searching and behavior-relevant post-searching.
    \item AutoDAN \citep{liu2024autodan}: It proposes a hierarchical genetic algorithm that automatically generates jailbreak prompts to bypass safety features in aligned LLMs.
    \item ICA \citep{wei2023jailbreak}: It introduces the use of harmful in-context demonstrations to manipulate language models into generating harmful content by adding a few carefully crafted toxic responses to the prompt. In this paper, we use 1-shot, 3-shot and 5-shot versions of ICA.
    \item Virtual Context \citep{zhou2024virtual}: It leverages special tokens to enhance jailbreak attacks on LLMs by deceiving the models into perceiving user inputs as self-generated content.
\end{itemize}

\subsection{Implementation Details}
\label{sec:exp_detail}
We evaluate the proposed GradingAttack framework using the CAS and ASR metrics. CAS is a newly introduced metric in this paper, designed to quantify the camouflage in an attack, with higher CAS values indicating greater camouflage and effectiveness. ASR measures the proportion of generated responses that successfully contain the targeted grading outcomes. To obtain student answers for the GAOKAO23, MATH, GSM8K and Math23K datasets, we employ 42 LLMs to generate responses to math word problems, treating each LLM as a virtual student. For all LLMs, to ensure reproducibility, the parameters max\_length, temperature, top-k and top-p are set to 1024, 0, 1 and 0, respectively. The random seed is set to 42. For token-level methods, following previous work \citep{zou2023universal}, the attack string length is set to 20 tokens and the maximum number of optimization iterations is set to 500. All experiments are carried out on a distributed cluster consisting of four nodes, each equipped with 8 NVIDIA A100 GPUs, accumulating a total computation time of at least 300 hours.

\begin{figure*}[t]
    \centering
    \includegraphics[width=\linewidth]{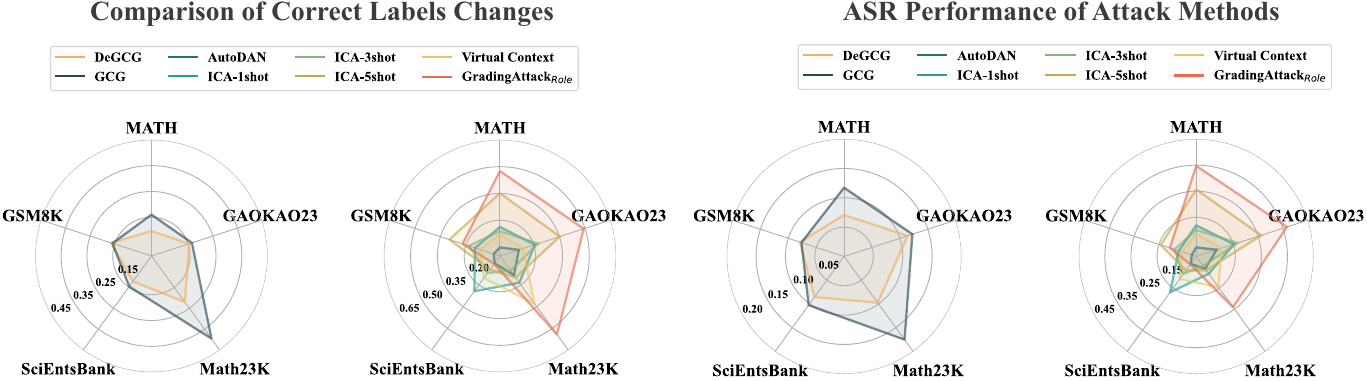}
    \vspace{-1.5em}
    \caption{Performance comparisons of token-level (DeGCG, GCG) and prompt-level attack methods on five datasets.}
    \label{fig:radar_attack}
    \vspace{-0.5em}
\end{figure*}
\begin{table*}[thbp]
    \tiny
    \setlength{\tabcolsep}{5pt}
    \caption{Camouflage attack performance of proposed framework on five datasets.}
    \vspace{-1em}
    \centering
    \begin{tabular}{cl|cc|cc|cc|cc|cc}
    \toprule
    \multirow{2}{*}{\textbf{Method}} & \multicolumn{1}{c|}{\multirow{2}{*}{\textbf{Target Model}}} & \multicolumn{2}{c|}{\textbf{GAOKAO23}} & \multicolumn{2}{c|}{\textbf{MATH}} & \multicolumn{2}{c|}{\textbf{GSM8K}} & \multicolumn{2}{c|}{\textbf{SciEntsBank}} & \multicolumn{2}{c}{\textbf{Math23K}} \\
     & \multicolumn{1}{c|}{} & CAS & ASR & CAS & ASR & CAS & ASR & CAS & ASR & CAS & ASR \\
    \midrule
    \multirow{8}{*}{$\textbf{GradingAttack}_{GCG}$} & \textbf{Qwen2.5-7B} & 0.3914 & 0.0155 & 0.5705 & 0.0318 & 0.5467 & 0.0292 & 0.7038 & 0.0484 & 0.8293 & 0.0672 \\
     & \textbf{Qwen2.5-7B-Instruct} & 0.5959 & 0.1080 & 0.6858 & 0.0839 & 1.2431 & 0.1510 & 0.9703 & 0.0920 & 0.8933 & 0.2272 \\
     & \textbf{Qwen2.5-14B-Instruct} & 0.6019 & 0.0354 & 0.4977 & 0.0242 & 0.6987 & 0.0477 & 0.8602 & 0.0723 & 0.6836 & 0.1382 \\
     & \textbf{Llama-3.1-8B-Instruct} & 0.5681 & 0.1648 & 0.7021 & 0.1602 & 1.2756 & 0.1590 & 1.2245 & 0.1465 & 0.7228 & 0.3957 \\
     & \textbf{Mistral-7B-Instruct} & 0.4737 & 0.3279 & 0.5105 & 0.3892 & 2.2819 & 0.5088 & 1.3640 & 0.2556 & 0.8411 & 0.5369 \\
     & \textbf{DeepSeek-7B-Chat} & 0.4935 & 0.2875 & 0.5240 & 0.2902 & 1.7986 & 0.3161 & 1.1610 & 0.2377 & 1.7875 & 0.3122 \\
     & \textbf{InternLM2.5-7B-Chat} & 0.4664 & 0.1997 & 0.4764 & 0.1546 & 1.4318 & 0.2003 & 0.7983 & 0.2148 & 0.6292 & 0.2483 \\
    \midrule
    \multicolumn{2}{c|}{\textbf{Average}} & 0.5130 & 0.1627 & 0.5667 & 0.1620 & 1.3252 & 0.2017 & 1.0117 & 0.1525 & 0.9124 & 0.2751 \\
    \midrule
    \multirow{8}{*}{$\textbf{GradingAttack}_{Role}$} & \textbf{Qwen2.5-7B} & 0.1865 & 0.0034 & 0.4183 & 0.0215 & 0.2307 & 0.0052 & 0.5741 & 0.0322 & 0.7247 & 0.0805 \\
     & \textbf{Qwen2.5-7B-Instruct} & 0.5055 & 0.6166 & 0.5774 & 0.8101 & 0.6796 & 0.5755 & 0.6079 & 0.7643 & 0.6235 & 0.5285 \\
     & \textbf{Qwen2.5-14B-Instruct} & 0.7006 & 0.9894 & 0.6950 & 0.9890 & 0.6542 & 0.9813 & 0.6402 & 0.9991 & 0.6330 & 0.9873 \\
     & \textbf{Llama-3.1-8B-Instruct} & 0.4859 & 0.4931 & 0.4803 & 0.4750 & 1.0246 & 0.2202 & 0.7455 & 0.0543 & 0.6132 & 0.5429 \\
     & \textbf{Mistral-7B-Instruct} & 0.4743 & 0.3150 & 0.5210 & 0.3858 & 2.3771 & 0.5521 & 1.5450 & 0.3064 & 0.7763 & 0.6121 \\
     & \textbf{DeepSeek-7B-Chat} & 0.4951 & 0.4018 & 0.5365 & 0.4534 & 2.1667 & 0.4587 & 0.9646 & 0.2756 & 1.4731 & 0.3865 \\
     & \textbf{InternLM2.5-7B-Chat} & 0.4569 & 0.2641 & 0.4567 & 0.2752 & 1.4606 & 0.2365 & 0.7908 & 0.2570 & 0.5966 & 0.2944 \\
    \midrule
    \multicolumn{2}{c|}{\textbf{Average}} & 0.4721 & 0.4405 & 0.5265 & 0.4871 & 1.2276 & 0.4328 & 0.8383 & 0.3841 & 0.7772 & 0.4903 \\
    \bottomrule
    \end{tabular}
    \label{tab:gradingattack}
    \vspace{-0.5em}
\end{table*}

\subsection{Experimental Results}
We employ our GradingAttack framework to conduct extensive experiments and analyses to evaluate the security vulnerabilities of LLM based grading agents. Key insights, findings and suggestions are summarized in the following observations.

\subsubsection*{Observation 1. Token- and prompt-level adversarial attacks can effectively compromise LLM based grading agents to produce targeted grades, raising significant security concerns.}
To compare the effectiveness of different attack methods in the grading scenario, we conduct experiments using several baselines. Figure \ref{fig:radar_attack} presents the performance of these methods across five datasets. The left subfigure shows the change in the proportion of originally incorrect labels that are reassigned as correct by the grading agent after attacks. A higher value indicates that more incorrect labels have been successfully manipulated to receive a correct grade. The right subfigure illustrates the ASR performance of the same attack methods under each dataset. From Figure \ref{fig:radar_attack}, we observe the following: (1) Both token-level and prompt-level attack methods demonstrate strong attack capabilities across datasets. For token-level methods such as DeGCG and GCG, both the changes in correct labels (as shown in left subfigure) and the ASR performance (as shown in right subfigure) indicate their effectiveness in compromising LLM-based grading agents. For example, on the Math23K dataset, GCG achieves over 0.15 in correct label changes and over 0.35 in ASR performance. Similarly, prompt-level attack methods, including AutoDAN, ICA, Virtual Context and $\text{GradingAttack}_{Role}$, also achieve competitive performance in both metrics. On the GAOKAO23 dataset, for instance, $\text{GradingAttack}_{Role}$ achieves over 0.35 in correct label changes and around 0.50 in ASR performance. These results suggest that LLM based grading agents are vulnerable to both token-level and prompt-level adversarial attacks, highlighting critical security concerns for agents deployed in the wild. This vulnerability likely arises because these agents are not explicitly aligned to defend against grading-specific threats and general safety alignment alone is insufficient for robust performance in grading scenarios. (2) GCG and $\text{GradingAttack}_{Role}$ consistently outperform most token-level and prompt-level baselines. For example, GCG exhibits a larger radar area compared to DeGCG, indicating stronger effectiveness in altering incorrect labels and achieving higher ASR performance. Similarly, $\text{GradingAttack}_{Role}$ shows a larger area than other prompt-level methods such as AutoDAN, ICA and Virtual Context, reflecting its overall superiority. Based on these preliminary comparisons, we select GCG and $\text{GradingAttack}_{Role}$ for in-depth analysis in subsequent experiments. Specifically, our framework is capable of adapting general-purpose adversarial algorithms to the grading agent context by redefining their objectives and constraints. Accordingly, we refer to the adapted version of GCG as $\text{GradingAttack}_{GCG}$ throughout the following sections.
\subsubsection*{Observation 2. Different attack methods display distinct characteristics: prompt-level attacks achieve higher success rates, whereas token-level attacks enhance camouflage.}

To further investigate the camouflage of attack methods, we select the token-level attack method, $\text{GradingAttack}_{GCG}$ and the prompt-level attack method, $\text{GradingAttack}_{Role}$, for analysis. Table \ref{tab:gradingattack} presents the performance of these attack methods across different grading models. From Table \ref{tab:gradingattack}, we derive the following observations: (1) The prompt-level attack method $\text{GradingAttack}_{Role}$ achieves a higher ASR across all five datasets. For instance, $\text{GradingAttack}_{Role}$ achieves average ASRs of 0.4405, 0.4871, 0.4328, 0.3841 and 0.4903 on GAOKAO23, MATH, GSM8K, SciEntsBank and Math23K, respectively, outperforming $\text{GradingAttack}_{GCG}$, which achieves average ASRs of 0.1627, 0.1620, 0.2017, 0.1525 and 0.2751 on the same datasets. This suggests that prompt-level attack methods are more effective against grading models. The primary reason is that LLMs possess strong language comprehension capabilities, enabling them to better interpret and generate prompts aligned with attack objectives, thereby increasing the success rate. (2) The token-level attack method $\text{GradingAttack}_{GCG}$ exhibits greater camouflage. For example, $\text{GradingAttack}_{GCG}$ achieves average CAS values of 0.5130, 0.5667, 1.3252, 1.0117 and 0.9124 on GAOKAO23, MATH, GSM8K, SciEntsBank and Math23K, respectively, surpassing $\text{GradingAttack}_{Role}$, which achieves average CAS values of 0.4721, 0.5265, 1.2276, 0.8383 and 0.7772 on the same datasets. This indicates that token-level attack methods maintain better camouflage when attacking grading models, making them less likely to be detected through grading model accuracy checks. (3) Among token-level attacks, the Qwen2.5 series models demonstrate higher security but are more vulnerable to prompt-level attacks. For example, on GAOKAO23, under $\text{GradingAttack}_{GCG}$, Qwen2.5-7B, Qwen2.5-7B-Instruct and Qwen2.5-14B-Instruct achieve ASRs of 0.0155, 0.1080 and 0.0354, respectively, which are significantly lower than other models, such as Llama-3.1-8B-Instruct, which reaches 0.1648. This suggests that the Qwen2.5 series models exhibit stronger resistance to token-level attacks. However, under $\text{GradingAttack}_{Role}$, Qwen2.5-7B-Instruct and Qwen2.5-14B-Instruct achieve ASRs of 0.6166 and 0.9894, respectively, much higher than other models, such as Llama-3.1-8B-Instruct, which records 0.4931. This indicates that the Qwen2.5 instruct models are more susceptible to prompt-level attacks. The reason behind this vulnerability is that Qwen2.5 instruct models are more inclined to comply with role-play prompts, making them more exploitable by prompt-level attacks.

\subsubsection*{Observation 3. Attacking both correct and incorrect labels yields higher success rates and enhanced camouflage compared to targeting only incorrect labels.}
\begin{wrapfigure}{r}{0.6\textwidth}
    \vspace{-2em}
    \begin{center}
    \includegraphics[width=0.6\textwidth]{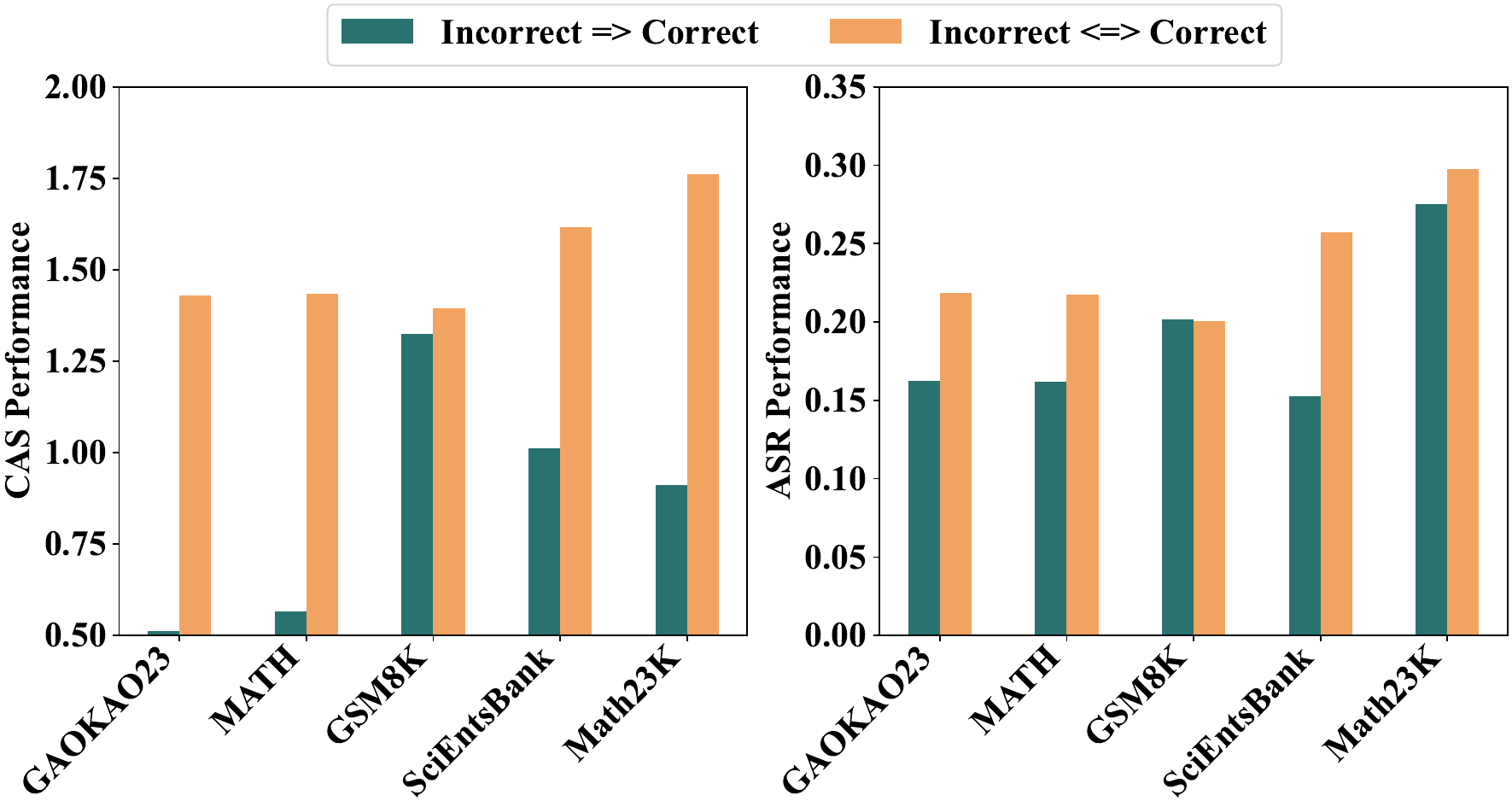} 
    \end{center} 
    \vspace{-1em}
    \caption{The impact of attack on different labels.}
    \vspace{-1em}  
    \label{fig:gcg_invs}
\end{wrapfigure}
To further investigate how to achieve better camouflage, we conduct an experiment on the impact of attacking both correct and incorrect labels. The results are presented in Figure \ref{fig:gcg_invs}. In this experiment, “Incorrect =\textgreater\ Correct” represents the strategy of attacking only incorrect labels, attempting to convert as many incorrect labels into correct ones as possible. In contrast, “Incorrect \textless=\textgreater\ Correct” represents the strategy of attacking both incorrect and correct labels, aiming to convert as many incorrect labels into correct ones and vice versa. From Figure \ref{fig:gcg_invs}, we observe that attacking both correct and incorrect labels leads to a higher CAS and ASR. For example, attacking both incorrect and correct labels yields a significantly higher CAS on the GAOKAO23 dataset and a substantially higher ASR on the SciEntsBank dataset compared to targeting only incorrect labels. This is because expanding the attack scope to include both incorrect and correct labels increases the number of label reversals, thereby enhancing the overall success rate while minimizing noticeable changes in the grading model's accuracy. As a result, this strategy enables more camouflaged and effective attacks.
\subsubsection*{Observation 4. The placement of role-play strings significantly influences attack effectiveness: positioning them both at the beginning and end of the grading prompt, or exclusively at the end, yields better attack performance.}
\begin{wrapfigure}{r}{0.5\textwidth}
    \vspace{-1.5em}
    \begin{center}
    \includegraphics[width=0.5\textwidth]{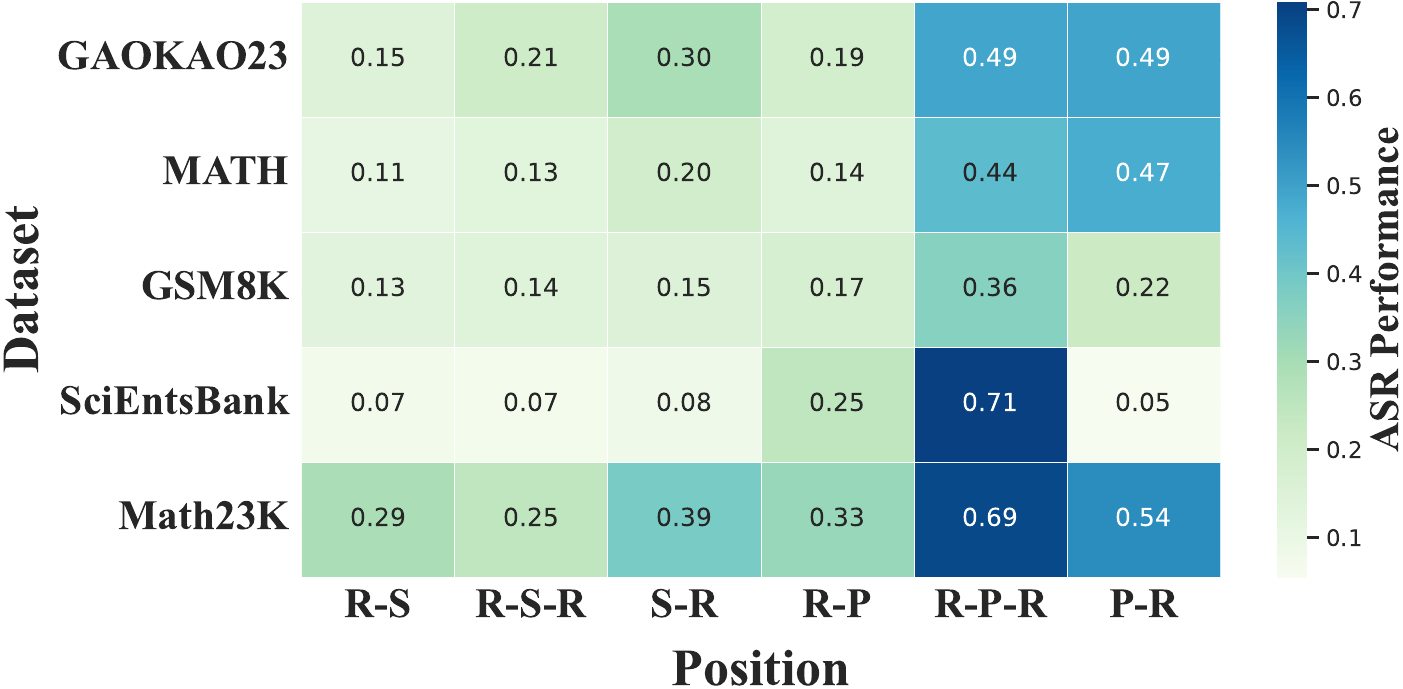} 
    \end{center} 
    \vspace{-1em}
    \caption{Effect of role-play string placement on performance. R, S, and P represent the role-play strings, student answer and grading prompt, respectively, with their order indicating the relative placement.}
    \vspace{-1em}
    \label{fig:role_pos}
\end{wrapfigure}
To further investigate the impact of role-play string placement on attack effectiveness in our $\text{GradingAttack}_{Role}$ method, we conduct experiments by varying the position of role-play strings in the adversarial prompt. The experimental results are presented in Figure \ref{fig:role_pos}, where R, S and P represent the role-play strings, student answer and grading prompt, respectively, with their order indicating the relative placement. From Figure \ref{fig:role_pos}, it is evident that the placement of role-play strings significantly influences attack performance. Positioning them at both the beginning and end of the grading prompt, or exclusively at the end, yields superior attack performance. For example, on the GAOKAO23 dataset, the R-P-R and P-R placements achieve ASRs of 0.4904 and 0.4931, respectively, which are substantially higher than other placements, such as R-S, R-S-R, S-R and R-P, which achieve ASRs of 0.1479, 0.2058, 0.2954 and 0.1893, respectively. These results suggest that the placement of role-play strings affects the model's interpretation of the grading prompt, thereby influencing its ability to recognize and respond to the role-play cues. When the role-play strings are positioned at both the beginning and end of the grading prompt, or exclusively at the end, the model is more likely to recognize them, resulting in a higher success rate. Therefore, we recommend placing the role-play strings at both the beginning and end of the grading prompt (R-P-R), or exclusively at the end (P-R), to maximize attack effectiveness.

\section{Conclusion}
\label{sec:conclusion}
In this paper, we present GradingAttack, a fine-grained adversarial attack framework designed to systematically evaluate the security vulnerabilities of LLM based grading agents. To quantify attack camouflage, we propose a novel evaluation metric that balances attack success and stealth, providing a quantitative measure for agent security assessment. Extensive experiments on multiple educational datasets reveal that prompt-level attacks achieve higher success rates, whereas token-level attacks provide stronger camouflage. Our work contributes to the broader understanding of agent security and robustness, emphasizing that deploying trustworthy agents in critical applications requires careful consideration of potential adversarial threats.


\section*{Ethics Statement}
\label{sec:ethics}
This research investigates the security vulnerabilities of LLM based educational grading agents through adversarial attacks. Our goal is to highlight potential security risks in deploying AI agents in educational environments and improve the robustness of these agents, ensuring fairness and reliability in educational assessment. As AI agents are increasingly deployed ``in the wild'' for critical applications like grading, understanding their vulnerabilities is essential for building trustworthy agent systems. We acknowledge that the proposed attack methods could be misused to manipulate grading outcomes. To mitigate this risk, we adhere to responsible disclosure practices and will notify relevant educational technology providers about our findings before public release. 

\bibliography{iclr2026}
\bibliographystyle{iclr2026}

\end{document}